\documentclass{emulateapj}

\usepackage{graphicx}
\usepackage{apjfonts}
\usepackage{mathptmx}



\def\***#1{{\sc #1}}
\def\plan#1{\relax}
\def\Plan#1{\relax}
\def\PLAN#1{\relax}

\def\lta{\mathrel{\spose{\lower 3pt\hbox{$\mathchar"218$}}
     \raise 2.0pt\hbox{$\mathchar"13C$}}}
\def\gta{\mathrel{\spose{\lower 3pt\hbox{$\mathchar"218$}}
     \raise 2.0pt\hbox{$\mathchar"13E$}}}
\newcommand{\etal}{{\it et al. }}

\def\mathnew{\mathsurround=0pt}

\def\simov#1#2{\lower .5pt\vbox{\baselineskip0pt \lineskip-.5pt
\ialign{$\mathnew#1\hfil##\hfil$\crcr#2\crcr\sim\crcr}}}

\def\simless{\mathrel{\mathpalette\simov <}}

\begin{document}

\title{Interpretation of the  Radio/X-ray Knots of 
AGN Jets within the Internal Shock Model Framework}

\author{S. Sahayanathan\altaffilmark{1} and R. Misra\altaffilmark{2} }

\altaffiltext{1}{Nuclear Research Laboratory, Bhabha Atomic Research Center, Mumbai, India; \newline sunder@apsara.barc.ernet.in}

\altaffiltext{2}{Inter-University Center for Astronomy and Astrophysics, Post Bag 4,
Ganeshkhind, Pune-411007, India; rmisra@iucaa.ernet.in}

\begin{abstract}
The dynamics of 
relativistically moving blobs ejected out of a central AGN, are 
considered. It is assumed
that the collision between two blobs are completely inelastic, such that
the bulk kinetic energy lost in the collision is used to energize electrons
to relativistic energies via acceleration in internal shocks which
are formed due to the collision. These high energy electrons
which are produced on a time-scale corresponding to the collision 
time-scale, cool by radiative losses due to 
synchrotron and Inverse Compton processes.
The model is applied to the radio/X-ray knots of several AGN.
For three of these sources we have analyzed long ($ > 40$ ksec)
{\it Chandra} observations and report on constrains on the
X-ray spectral indices. 
In the framework of this model the AGN are
inferred to sporadically eject relativistic blobs on time-scales
ranging from $10^{11}$ to $10^{12}$ secs for different sources.
It is shown that the collision time-scales can be longer
than the age of the knot and, hence, non-thermal electrons are
continuously being injected into the system. This continuous
injection, in contrast to an instantaneous one time injection, gives
rise to a characteristic spectral break, rather than a high-energy
cutoff in the spectrum.

\end{abstract}


\section{Introduction} 

Study of the knots in kilo parsec scale jets of Active Galactic Nuclei (AGN) 
has been given a new dimension 
after the advent of {\it{Chandra}} which due to its excellent 
spatial resolution 
is able to resolve bright X-ray knots of such jets \citep{chartas, samb02, samb01,tav}. In most of the cases, the
X-ray knots coincide with their radio/optical counterparts 
\citep{samb02,pesce,samb01,marsh,wilson}. The discovery of such knots 
in different energy bands 
provides useful information about the basic emission mechanism and the 
underlying acceleration processes involved there. The 
radio and optical emission 
from these knots are generally accepted to be of synchrotron origin, whereas the
X-ray emission can be due to either synchrotron 
\citep{worall,samb02,pesce,sahay} or Inverse Compton processes 
\citep{samb01,samb02,pesce,tav,schw,chartas,sahay}. 
If the radio-to-optical index ($\alpha_{RO}$) is larger than the 
optical-to-X-ray
index ($\alpha_{OX}$), a single emission mechanism cannot explain the observed 
fluxes (however see \cite{Der02} who show that if Klein-Nishina   losses 
is important, spectral hardening can occur) and the X-ray emission may be 
due to Inverse Comptonization. On the 
other hand, if $\alpha_{RO}$ is smaller than $\alpha_{OX}$, synchrotron origin 
of X-ray is acceptable \citep{tav,samb02,pesce}. This synchrotron
origin of X-rays for knots with $\alpha_{RO} < \alpha_{OX}$, was strengthened
for the knots of $3C 271$ by \cite{pesce} who show that the alternate
Inverse Compton model would require 
exceptionally large Doppler factors.
When the X-ray emission can be attributed to the Inverse Compton process, 
the possible 
choices of target photons are radio/optical synchrotron photons (SSC)\citep{schw} or the 
cosmic microwave background (IC/CMB)\citep{tav,samb01,samb02,pesce,sahay}. \cite{tav} have shown that the SSC 
interpretation would require large jet powers and magnetic fields much lower
than the equipartition values whereas IC/CMB requires relatively low 
jet power and near equipartition magnetic fields.

These possible radiative process identifications have to be
associated with (and confirmed by) dynamical models regarding the
origin and subsequent evolution of the radiating non-thermal particles.
In many models, these non-thermal particles are assumed to be 
generated by a short duration acceleration process and the particle
distribution is determined by radiative losses
\citep{samb01,samb02,tav,jaffe,karda,pach}. The high energy 
particles cool more efficiently and hence get depleted in time,
giving rise to a time-dependent 
high energy cut off in the non-thermal particle distribution.
If the X-ray emission is attributed to synchrotron emission
by these particles, then these models predict an
exponentially decreasing X-ray spectrum \citep{samb01,samb02,tav,pesce}.
On the other hand, it may also be possible that the acceleration process
exists for longer than the age of the knot, and hence, there is a continuous
injection of non-thermal particles. 
In this case, a time dependent break in the non-thermal 
particle distribution occurs  at $\gamma = \gamma_{c}$, where
$\gamma_c$ is determined by the condition that the cooling time-scale for
electrons with $\gamma = \gamma_{c}$ is equal to the age of the knot
\citep{heavens,meis}. This model predicts the X-ray spectral index 
$\alpha_{X}$ to be $\approx \alpha_{R}+1/2$, where $\alpha_{R}$ is the radio
spectral index. Thus the predicted spectrum depends, in particular, on
the duration of the non-thermal particle production and, in general, on the
production mechanism. 

While there is no consensus on the
origin of these non-thermal particles, one of the standard models
is the internal shock scenario \citep{rees78,spa}, where the
particles are energized by Fermi acceleration in shocks produced
during the interaction of relativistically moving blobs ejected
from the central engines with different speeds. This model has
also been used to explain the prompt emission of Gamma-ray Bursts
\citep{rees94,lazzati}. A detailed description of the shock formation
and subsequent electron acceleration is complicated and would require
numerically difficult magneto-hydrodynamic simulations.
Moreover, the limited number of observables, which can be obtained
from the featureless spectra in two or three different energy bands,
may not be able to constrain the various assumption and/or the
initial conditions of such a detailed study. Nevertheless, 
a qualitative idea as to whether the internal shock model is
consistent with the present observations and if so, qualitative
estimates of the model parameters would be desirable. Such an
estimate will provide insight into the temporal behavior of
the central engine.

In this work, we implement an internal shock model with simplifying
assumptions and compute the time-evolution of the non-thermal particles
produced. We compare the results obtained with the broad-band fluxes
from knots of several AGN jets and their observed positions. The motivation
here is to find a consistent set of model parameters that can explain the
observations and thereby make qualitative estimates of their values.
Apart from the fluxes at different energy bands, the spectral indices
in each band can also provide important diagnostic information about
the nature of these sources. Hence, we have
analyzed long ($ > 40$ ksec)
{\it Chandra} observations of three AGN and report the constrains that
were obtained on the
X-ray spectral indices of the individual knots.

In the next section, a brief description of the data analysis technique and
the results obtained are presented. In \S 3, 
the internal shock model and the assumptions
used in this work are described while  in \S 4, the results of the
analysis are presented and discussed.
Throughout this work, $H_o = 75$ km s$^{-1}$ Mpc$^{-1}$ and 
$q_0 = 0.5$ are adopted.

\section{Data Analysis}
Long exposure {\it{Chandra}} observations of the sources 
$1136-135$, $1150+497$
and $3C 371$ were performed with the ACIS-S with the source at the aim-point of the
S3 chip. The observation ID (Obs. ID) and the exposure time of the
observation are tabulated in Table 1. 
Earlier shorter duration observations of these sources had revealed
two bright knots for each source whose positions from the nucleus are
tabulated in Table 2. These longer duration observations allow for
better constrain on the X-ray spectral indices of these knots.

\begin{deluxetable} {rcccccccc}
\tablewidth{0pt}
\tablecaption{{\it{Chandra}} Observations}
\tablehead{
\colhead{Source name} & \colhead{Obs Id}  & \colhead{Exposure} & \colhead{Knots} &
\colhead{$F_{0.3-3.0}$} & \colhead{$\alpha_X$}}
\startdata
 $1136-135$  & $3973$ & $77.37$ & A & $0.63$ & $1.24^{+1.51}_{-0.66}$ \\ 
   & & & B &  $1.56$ & $0.65^{+0.69}_{-0.28}$\\  
  & & & & & \\
 $1150+497$ & $3974$ & $68.50$ & A & $3.15$ & $0.66^{+0.28}_{-0.26}$  \\
  & & & B & $0.61$ & $0.90^{+2.22}_{-1.02}$  \\
  & & & & & \\
 $3C 371$  & $2959$ & $40.86$ & A & $3.32$ & $1.43^{+0.85}_{-0.76}$  \\
   & & & B & $7.84$ & $1.07^{+0.26}_{-0.23}$  \\
\enddata
\tablecomments{Columns:- 1: Source name.  2: {\it Chandra} 
Observation Id 3: Exposure time (in ksec) 4: Knots prominent in X-ray 
5: Flux in $0.3-3.0$ keV energy band in  
$ergs/cm^2/s$. 6: X-ray energy spectral index.}
\end{deluxetable}

The reprocessed data from {\it{Chandra}} X-ray center were analyzed using 
the latest 
calibration files to produce the image and spectra. The X-ray
counts from each individual knot was extracted using a circular region
centered at the knot. 
The background was estimated from the counts obtained from 
same size regions located at the same 
distance from the nucleus but at  different 
azimuth angles. The radius of the circular region was chosen to be 
$0.74\arcsec$
for the sources $1136-135$ and $1150+497$, while for $3C 371$
a smaller radius of $0.6\arcsec$ was used. These sizes were chosen
to minimize any possible contamination from the nucleus and/or the
other knot.

\begin{deluxetable} {rcccccccc}
\tablewidth{0pt}
\tablecaption{Observed Knot Features}
\tablehead{
\colhead{Source name} & \colhead{Type} & \colhead{z} & \colhead{Knot} & \colhead{Position} & \colhead{$\alpha_{RO}$} & \colhead{$\alpha_{OX}$} & \colhead{Ref}}
\startdata
 $1136-135$ & FSRQ & $0.554$ & A & $4.5$ & $0.73$ & $0.83$ & 1\\ 
 &  &  & B & $6.7$ & $1.04$ & $0.68$ &\\ 
 & & & & & & & \\
 $1150+497$& FSRQ & $0.334$ & A & $2.1$ & $0.99$ & $0.83$ & 1 \\ 
 &  &  & B & $4.3$ & $1.24$ & $0.44$ & \\ 
 & & & & & & & \\
 $1354+195$& FSRQ & $0.720$ & A & $1.7$ & $1.05$ & $0.6$ & 1\\ 
 &  &  & B & $3.6$ & $1.15$ & $0.68$ & \\ 
 & & & & & & & \\
 $3C 273$& QSO & $0.158$ & A & $13$ & $0.86$ & $0.61$ & 2\\ 
 &  &  & B & $15$ & $0.9$ & $0.73$ & \\ 
 & & & & & & & \\
 $3C 371$& Bl Lac & $0.051$ & A & $1.7$ & $0.9$ & $1.28$ & 3\\
  &  &  & B & $3.1$ & $0.76$ & $1.14$ & \\ 
\enddata
\tablecomments{Columns:- 1: Source name. 2: Type of the source. 3: Redshift.
4: Knots prominent in X-ray. The nomenclature is same for all the knots as they are 
in the literature except for $3C 371$ where A and B are reverse. 5: Position of the 
knot in arc seconds. 6: Radio-to-Optical index. 7: Optical-to-X-ray index. 8: References: 1: \cite{samb02}, 2: \cite{samb01}, 3: \cite{pesce}}
\end{deluxetable}

Spectral fits were undertaken on the data 
using the XSPEC package in C statistic mode which is the appropriate
statistic when the total counts are low. The flux and the energy
spectral indices obtained are tabulated in Table 1.

\section {The Internal Shock Model}

In the internal shock model framework, temporal variations of the
ejection process produces density fluctuations (moving with
different velocities) which collide at some distance from
the source to produce an observable knot. This distance will
depend on the time-scale over which the variation takes place.
In general the system will
exhibit variations over a wide range of time-scales and
knots like features would be produced at different
distance scales. In this work, we consider large scale jets
(with deprojected distances $\approx 100$ kpc) which are
expected to arise from variability occurring on a corresponding large 
time-scale. Variations
on smaller time-scales would produce knot structures on 
smaller distance scales, for example  par-sec scale jets or even smaller,
which will be unresolved 
for these sources.
These smaller time-scale variabilities will be smoothened out
at large distances and hence one expects that 
the jet structure of these sources to be determined by 
variations over a single characteristic time-scale. To further simplify
the model, we approximate the density and velocity fluctuations
as two discrete 
blobs with equal masses, $M_1 = M_2 = M$, 
having Lorentz factors,
$\Gamma_1$ and $\Gamma_2$, that are ejected one after the other,
from the central engine with a time delay of $\Delta t_{12}$. The
collision of the blobs is considered to be completely inelastic i.e. 
the blobs coalesce and
move as a single cloud, which is identified with the observed knot. 
From conservation of momentum, the Lorentz factor of the knot is
\begin{equation}
\Gamma = \sqrt{(\frac {\Gamma_1 \beta_1 +\Gamma_2 \beta_2 }{2})^2+1}
\end{equation}
where $\beta_{1,2}=v_{1,2}/c$, are the velocities of the blobs normalized
to the speed of light.
Since the collision is inelastic a fraction of the bulk kinetic energy
is dissipated.
Denoting all quantities in the rest frame of the knot by subscript $K$,
this energy $\Delta E_{K}$ can be estimated to be,
\begin{equation}
\Delta E_K = [\Gamma_{1K} + \Gamma_{2K}-2)] M c^2
\end{equation}
The Lorentz factors of the blobs in the knot's rest frame 
$\Gamma_{1K,2K}=(1-\beta^2_{1K,2K})^{-1/2}$ are computed using
\begin{equation}
\beta_{1K,2K} = \frac {\beta_{1,2}-\beta}{1-\beta_{1,2}\beta}
\end{equation}
where $\beta = \sqrt{1-1/\Gamma^2}$.
The time-scale on which this energy will be dissipated may
be approximated to be the crossing-over time of the two blobs,
\begin{equation}
T_{ON,K} \approx \frac {2 \Delta x_{K}}{c(\beta_{2K}-\beta_{1K})}
\label{eqn:TON}
\end{equation}
where, $\Delta x_{K}$ is the average size of the two blobs in the rest frame
of the knot.  
It is assumed that  this dissipated bulk 
kinetic energy, $\Delta E_K$, gets converted efficiently  to the energy
of the non-thermal particles produced during the collision. 
The number of non-thermal particles injected per unit
time into the knot is taken to be,
\begin{equation}
Q_K (\gamma) d\gamma = A \gamma^{-p} d\gamma \;\;\; \hbox  {for}\;\;\; 
\gamma > \gamma_{min}{;} 
\end{equation}
where $\gamma$ is the Lorentz factor of the electrons, 
$p$ is the particle index and $A$ is the normalization constant 
given by,
\begin{equation}
A = \frac{\Delta E_K}{T_{ON,K}}\frac{(p-2)}{mc^2}\gamma_{min}^{(p-2)} 
\end{equation}
Here the injection is assumed to be uniformly occurring for a
time $T_{ON,K}$.  The cloud is permeated with a tangled 
magnetic field $B_K$. 

The kinetic equation describing the evolution of the
total number of non-thermal particles in the system, $N(\gamma,t_K)$, is
\begin{equation}
{\partial N_K(\gamma,t_K) \over \partial t_K} + {\partial  \over \partial \gamma} [P(\gamma,t_K) N_K(\gamma,t_K)] = Q_K (\gamma)
\label{evol}
\end{equation}
$P(\gamma,t)$ is the cooling rate given by
\begin{equation}
P(\gamma,t_K) = -(\dot \gamma_{S} (t_K) + \dot \gamma_{IC} (t_K) )
\end{equation}
where $\dot \gamma_{S} (t_K)$ and $\dot \gamma_{IC} (t_K)$ are
the cooling rates due to synchrotron and Inverse Compton of the CMBR 
respectively \citep{sahay}.

The non-thermal particle distribution $N_K(\gamma,t_K)$ produces
the  synchrotron
and Inverse Compton spectra which are computed at an observation time 
$t_K = t_{K,O}$. The Inverse Compton spectrum is computed after taking
into account the anisotropy of the CMBR spectrum in the rest frame
of the plasma \citep{der}.
Finally, the flux is transformed from the 
source frame to the observer's frame in earth taking care of the 
Doppler boosting \citep{begel} and cosmological effects.

Eqn (\ref{evol}) can be solved  analytically \citep{kirk} or
numerically computed using the technique given in
\cite{chang}. Rather than the complete analytical
expression it is perhaps more illuminating to study the asymptotic limits.
It is convenient to define a critical lorentz factor
$\gamma_c$ for which the cooling time-scale for synchrotron and 
IC losses is equal to the observation time $t_{K,O}$.
Then for $\gamma << \gamma_c$, there is effectively no cooling and
the particle spectrum is $N (\gamma, t) \propto \gamma^{-p} t_{K,O}$ i.e.
the injection rate times the observed time. For $\gamma >> \gamma_c$
the cooling time is shorter than the observed time and hence the
particle spectrum is in quasi-equilibrium 
$N (\gamma, t) \propto \gamma^{(p-1)/2}$. These two regimes in the
particle spectra give rise to a composite synchrotron spectra with
a spectral break.

Adiabatic cooling has been neglected in Eqn (\ref{evol}). 
The adiabatic cooling 
time-scale, $t_{adb}$ is $\approx R(t)/v_{exp}$ where $R(t)$ is 
the size of the knot and $v_{exp}$ is the speed at which the blob 
is expanding. Since $R(t_{K,O}) = R_i + v_{exp}t_{K,O}$, where $R_i$ is the 
initial size of the knot, it follows that $t_{adb}$ is always
greater than  $t_{obs}$ for the case when the initial size of
the blob $R_i \approx R(t_{K,O})$ and hence adiabatic cooling can be 
neglected. In the other extreme, if $R_i << R_ (t_{K,O})$,
$t_{abs}$ is at at most $t_{obs}$. For $\gamma >> \gamma_c$, 
the cooling time-scale is much smaller than $t_{obs} \approx t_{adb}$ and
hence adiabatic cooling may still be neglected.  For $\gamma << \gamma_c$,
since $t_{K,O} \approx t_{adb}$ and not much larger, the effect of adiabatic
cooling can only change the number density by a factor of few. Thus
for an order of magnitude calculation as required for this work, 
adiabatic cooling can always be neglected.

The predicted spectrum and size of a knot depends on 
nine parameters, which are
the mass of the blobs $M$, their average size $\Delta x_{K}$, the
Lorentz factors $\Gamma_1$ and $\Gamma_2$, the particle injection injection index
$p$, the minimum Lorentz factor $\gamma_{min}$, the initial 
magnetic field $B_0$,
the inclination angle of the jet $\theta$ and the observation time $t_{K,O}$.

From these parameters and the location of the knot in the sky plane, one
can infer the time delay $\Delta t_{12}$ between the ejection of the two blobs.
The projected distance of the knot from the source $S$ can be written as
\begin{equation}
S =  c ( \beta_1 t_c + \beta t_{O})\; \hbox {sin}\theta
\label{Seqn}
\end{equation}
where $t_{O} = \Gamma t_{K,O}$ is the time of the observation after
the formation of the knot in the source frame. The time elapsed $t_c$, 
after the ejection of the first blob and the start of the collision,
is given by,
\begin{equation}
t_c = \frac{v_2 \Delta t_{12} - \Delta x_1}{v_2 - v_1}
\end{equation}
where $\Delta x_1$ is the size of the first blob 
$\approx \Gamma_{1,K} \Delta x_{K}/\Gamma$.
Thus $\Delta t_{12}$, can be estimated using the above equation where $t_c$ is
given by Eqn (\ref{Seqn}), and it essentially depends on 
four parameters, $\theta$, $\Gamma_1$, $\Gamma_2$ and $t_{K,O}$.

A total time $t_{tot}$ can be defined to be the time that has elapsed
between the ejection of the first blob and the observation, $t_{tot} = 
t_c + t_{O}$. 
For two knots, A and B, the time difference between the ejection of their 
first blobs, $t^{AB}$ is then
\begin{equation}
t^{AB} = t^A_{tot} - t^B_{tot} - t_{LT}
\end{equation}
where $t_{LT}$ is the light travel time difference between the two knots
which is approximated to be
\begin{equation}
t_{LT} = \frac{S^A - S^B}{c \;\;\hbox {tan} \theta} 
\end{equation}

The power of the jet, can be defined in two different ways. The instantaneous
power, which is the power when the system is active, can be defined to
be the average energy of the blobs divided by the time-scale on which
the blobs are ejected. This power can be 
estimated for each knot to be
\begin{equation}
P_{ins} \approx \frac {M c^2 (\Gamma_1 + \Gamma_2)/2}{\Delta t_{12}}
\end{equation}
On the other hand,
the time averaged power of the jet can be defined as the 
typical energy ejected
during active periods divided by the time-scale on which such activity
occur. For two knots, A and B, this can be approximated to be
\begin{equation}
P_{ave} \approx \frac {[M^A c^2 (\Gamma^A_1 + \Gamma^A_2) + M^B c^2 (\Gamma^B_1 + \Gamma^B_2)]/2}{\Delta t^{AB}}
\end{equation}

Two {\it a posteriori} checks have to be imposed to ensure self-consistency.
The total number of non-thermal particles that is injected, $N_{nth}$ should
be less than the number of particles in the knot, $N_K \approx 2 M/m_p$ and
that the magnetic field $B_K (t_K)$ should be less than the equipartition
value, $B_{equ}$. 

\section{Results and Discussion}

The model has been applied to those knots of kpc-scale jets, which have
been detected by {\it Chandra}, and for which radio and optical data are available.
This criterion was satisfied by the two brightest knots of the AGN:
 $1136-135$, $1150+497$, $1354+195$, $3C 273$ and $3C371$. In this
work, the knot closer to the nucleus is referred to as Knot A and the
further one as Knot B. For these sources this nomenclature is same as 
in the literature except for $3C 371$ where the farther one has been
referred to as Knot A. For three of these sources, the X-ray spectral
indices were constrained using long exposure observations as described
in \S 2. The observed properties of the sources and the knots 
are tabulated in Table $1$ and $2$.

Figure $1$ shows the observed radio, optical and X-ray fluxes of these knots,
along with the computed spectra corresponding to model parameters that  
are given in Table $3$. Since the number of parameters is large as compared
to the observables, a unique set of parameter values cannot be obtained.
Two consistency checks have been imposed on the parameter values, which
are that the number of non-thermal electrons which will be injected into
the system, $N_{nth}$, is smaller than the total number of protons, $N_k$,
and that the
magnetic field, $B$ is less than the equipartition value, $B_{equ}$. 
Both these conditions
are satisfied by the parameter sets as shown in Table $3$, where the
ratios $B/B_{equ}$ and $N_{nth}/N_K$ are given.

\begin{figure}
\begin{center}
{\includegraphics[width=1.0\linewidth,angle=0]{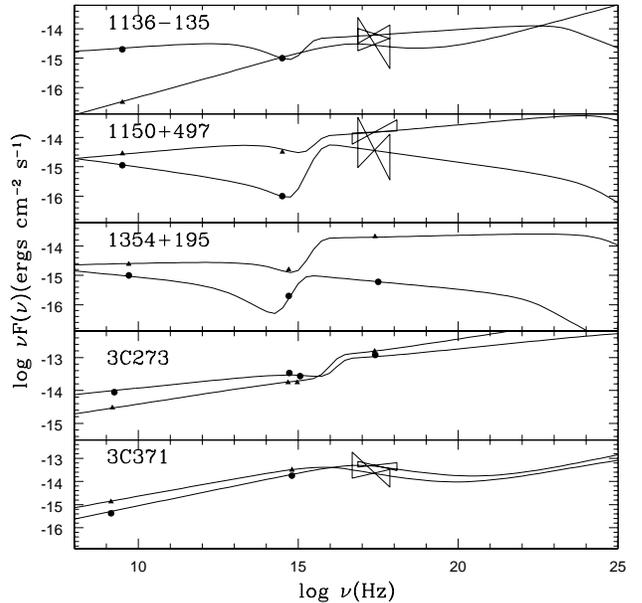}}
\end{center}
\caption{ The observed fluxes in radio, optical and X-ray compared
with model spectra using parameters given in Table 2. Knot A fluxes are represented 
by filled triangles and knot B fluxes are represented by filled circles. Errors
are typically 30\% or larger.\label{Figure 1}  }
\end{figure}

For each source, 
the time delay, $\Delta t_{12}$, between
the ejection of the two blobs which form the knots are
nearly equal to the time difference between the ejection of the first blobs
of Knot A and Knot B, $t^{AB}$. This gives an overall {\it single} time-scale of
activity for each source which ranges from $10^{11} - 10^{12}$ sec which
can reproduce the knot properties as had been assumed in the
development of the simple internal shock model. 
This result is important since if it had not been true  a more
complex temporal behavior would have to be proposed, wherein
the jet structure is due to variability of the source in two
different time-scales. The first being the time difference between
the ejection of two blobs that form a knot and the second being
the time difference between the activities that produced the two
knots.

The cross over
time, $T_{ON,K}$, is determined here by Eqn. \ref{eqn:TON}. Since,
$\beta_{2K}-\beta_{1K} \approx (\Gamma_1 - \Gamma_2)/\Gamma \approx 0.3$,
$T_{ON,K} \approx 10^{12} (\Delta x_K / 5 \times 10^{21} \hbox {cm})$ sec.
During this time, i.e. the time when there is injection of particles
into the system, the knot would travel a distance $\approx c T_{ON} 
\approx c \Gamma T_{ON,K} \approx 50 (\Gamma/5) (\Delta x_K / 5 \times 10^{21} \hbox {cm})$ kpc. This is a significant fraction of the total
observable distance traveled by the knot, from formation 
$\approx c t_c \approx 100$ kpc (see Table 3) to the termination
of the jet in the radio lobe, $\approx 200$ kpc. Hence, as
shown in Table 4, it is possible to fit the spectra of all
the knots, with an observation time, $t_{O,K}$ which is less than 
the crossing over time, $T_{ON,K}$ implying that there is  continuous 
injection of particles into the system.
In this scenario, the
synchrotron and IC spectra have a break corresponding to a Lorentz
factor, $\gamma_c$, where the cooling time-scale equals the 
observation time \citep{sahay}. This is in contrast to one-time
injection models ( i.e. when $t_{O,K} > T_{ON,K}$) where an
exponential cutoff in the spectra occurs. However, this
does not exclude the possibility that for some knots, 
$t_{O,K} > T_{ON,K}$ and such cutoff in spectra would be 
detected.

\begin{deluxetable*} {ccccccccccccc}
\tablewidth{0pt}
\tablecaption{Model Parameters}
\tablehead{
\colhead{Source} & \colhead{Knot}  & \colhead{$\theta$} & \colhead{$\Gamma_1$} & \colhead{$\Gamma_2$} & \colhead{$\Gamma^\ast$} & \colhead{$t_{KO}$} & \colhead{$t_c^\ast$} & \colhead{$t_{tot}^\ast$} & \colhead{log$\,$M} & 
\colhead{$\gamma_{min}$} & \colhead{p} & \colhead{$B$}}
\startdata
$1136-135$ & A & $11.5$ & $4.6$ & $5.4$ & $5.0$ & $0.25$ & $11.8$ &  $11.6$ & $38.0$ & $2$ & $2.4$ & $1.1$ \\ 
 & B &  & $4.1$ & $5.9$ & $5.0$ & $8.5$ & $13.1$ & $17.3$ & $36.4$ & $20$ & $2.9$ & $4.5$\\ 
$1150+497$ & A & $10.2$ & $2.5$ & $3.0$ & $2.8$ & $16.0$ & $1.2$ & $5.1$ & $37.5$ & $30$ & $2.8$ & $1.4$ \\ 
& B &  & $2.6$ & $3.5$ & $3.0$ & $6.2$ & $8.8$ & $10.0$ & $36.8$ & $30$ & $3.3$ & $4.0$\\ 
$1354+195$ & A & $8.21$ & $1.7$ & $2.3$ & $2.0$ & $9.0$ & $8.1$ & $8.1$ & $38.1$ & $37$ & $3.0$ & $1.7$\\ 
& B &  & $1.7$ & $2.3$ & $2.0$ & $8.7$ & $15.0$ & $17.0$ & $37.1$ & $20$ & $3.2$ & $8.0$\\ 
$3C 273$ & A & $8.23$ & $2.2$ & $4.0$ & $3.1$ & $2.1$ & $27.0$ & $24.0$ &  $36.8$ & $50$ & $2.7$ & $0.4$\\ 
& B &  & $1.7$ & $2.3$ & $2.0$ & $8.0$ & $30.0$ & $32.0$ & $37.8$ & $80$ & $2.8$ & $0.6$\\ 
$3C 371$ & A & $15.8$ & $1.6$ & $2.4$ & $2.0$ & $5.9$ & $0.18$ & $1.4$ &  $35.5$ & $20$ & $2.5$ & $1.0$\\ 
& B &  & $2.0$ & $2.4$ & $2.2$ & $2.3$ & $0.39$ & $0.75$ & $36.8$ & $10$ & $2.4$ & $0.6$\\ 
\enddata
\tablecomments{Seven of the model parameters and derived quantities. 
The eight parameter,
$\Delta x_K = 5.0 \times 10^{21}$ cm for all sources. Columns marked
with $\ast$ are derived quantities and not parameters.
Columns:- 1: Source name. 2: Knot. 3: Viewing angle (in degrees).
4: Lorentz factor of the first blob. 5: Lorentz factor of the second blob. 6: Lorentz factor of the Knot. 7: Observation time (in $10^{11}$ sec). 8: Collision time (in $10^{12}$ sec). 9: Total time (in $10^{12}$ sec). 10:  Mass of the blobs (in g). 11: Minimum Lorentz factor of the particle injected into the knot. 12: Injected particle spectral index. 13: Magnetic field (in $10^{-5}$G).} 
\end{deluxetable*}

The knots of 3C371 and 
Knot A of 1136-135 are unique in this sample since 
their the X-ray flux lies below the extrapolation of
the radio/optical spectra to X-ray wavelengths. This allows
for the interpretation that the X-ray flux is due to synchrotron
emission \citep{samb02,pesce,sahay}. For Knot A of 1136-135 and
Knot B of 3C371, this implies that the spectral break 
for the synchrotron emission occurs at the X-ray regime (Figure 1) 
which in
turn indicates that these sources are relatively younger ones. Indeed,
the ratio of the observation time to the time-scale of injection, 
$t_{O,K}/T_{ON,K}$ for these sources are smallest (Table 4). 
On the other hand, for Knot A of 3C371, the spectral break
can occur at the optical band even if the X-ray flux is interpreted
 as being due to synchrotron emission (Figure 1) and hence this source
need not be relatively young. However, this is only possible if
$t_{O,K} < T_{ON,K}$ and there is continuous injection
of particles. Otherwise, a sharp cutoff in the spectrum at the optical 
band would have occurred and the X-ray emission would not be due to
synchrotron emission.

\begin{deluxetable*} {ccccccccccc}
\tablewidth{0pt}
\tablecaption{Knot/Jet Properties}
\tablehead{
\colhead{Source} & \colhead{Knot} & \colhead{$\Delta t_{12}$} & \colhead{$t^{AB}$} & \colhead{$\frac{B}{B_{equ}}$} & \colhead{$\frac{N_{nth}}{N_K}$} & \colhead{log$\,P_{ins}$} & \colhead{log$\, P_{ave}$} & \colhead{$T_{ON,K}$} & \colhead{$\frac{t_{K,O}}{T_{ON,K}}$} & \colhead{D}}
\startdata
$1136-135$ & A & $1.1$ & $3.1$ & $0.55$ & $0.88$ & $48.60$ & $48.17$ & $20.4$ & $0.01$ & $112.8$ \\ 
  & B & $2.5$ & & $0.74$ & $0.74$ & $46.65$ & & $9.1$ & $0.93$ & $163.4$ \\ 
$1150+497$ & A & $0.9$ & $5.0$ & $0.12$ & $0.13$ & $47.98$ & $47.31$ & $17.0$ & $0.94$ & $50.5$ \\ 
 & B & $3.9$ & & $0.65$ & $0.43$ & $46.63$ & & $10.7$ & $0.58$ & $96.1$ \\ 
$1354+195$ & A & $7.5$ & $18.0$ & $0.04$ & $0.38$ & $47.49$ & $47.15$ & $9.6$ & $0.94$ & $78.7$ \\ 
 & B & $17.0$& & $0.64$ & $0.78$ & $46.10$ & & $9.6$ & $0.90$ & $135.5$ \\ 
$3C 273$ & A & $20.0$ & $34.0$ & $0.03$ & $0.80$ & $46.00$ & $46.63$ & $5.4$ & $0.39$ & $240.7$ \\ 
 & B & $31.6$ & & $0.02$ & $0.16$ & $46.60$ & & $9.6$ & $0.84$ & $248.5$\\ 
$3C 371$ & A & $1.4$ & $1.5$& $0.39$ & $0.88$ & $45.59$ & $46.98$ & $7.2$ & $0.83$ & $11.2$ \\ 
 & B & $1.0$ & & $0.28$ & $0.29$ & $47.14$ & & $16.3$ & $0.14$ & $7.7$ \\ 
\enddata
\tablecomments{Columns:- 1: Source name. 2: Knot. 3: Time delay between the ejection of the blobs (in $10^{11}$ sec). 4: Time delay between the ejection of the first and the third blob (in $10^{11}$ sec). 5: Ratio of the magnetic field to equipartition magnetic field. 6: Ratio of non-thermal electrons to the total number of electrons. 7: The instantaneous power (in ergs/sec). 8: Time averaged power (in ergs/sec). 9: Time-scale over which non-thermal particles are injected (in $10^{11}$ sec). 10: Ratio of the observation time to particle injection time-scale. 11: Deprojected distance in kilo parsec}
\end{deluxetable*}

Figure 1 shows that the radio, optical and X-ray spectral
indices for different knots may vary and highlights the
need for more spectral measurements in all bands. A definite
prediction of this model is that for most knots, the X-ray
spectral index should be equal to the radio one, indicating
that the X-ray flux is due IC/CMBR. Such spectral constrains would
be  particularly important since,
although it has been demonstrated here that the internal
shock model can explain the broad band spectra of these
sources, there could be other models which may be physically
and observationally more favorable. Recently \cite{Jes05}
have analyzed VLA and HST image of $3C 273$ and have found
that the optical and radio spectral indices are different
indicating the presence of an additional emission mechanism
for the source. Earlier,
\cite{ATO04}
argued that the X-ray flux is due to a second population of
non-thermal electrons, rather than being the IC/CMBR spectra
of the same distribution which produces the radio and optical
emission. They point out in the IC/CMBR model, since the X-ray emission
is due to electrons which are only a factor ten more energetic than 
those which produce the
radio, a source where 
the X-ray flux falls rapidly from the center should also exhibit
a similar decrease in radio emission which is not observed
( e.g. 3C273 and 1354+195). Moreover, the jet power required
in the IC/CMBR model can be very large $\approx 10^{48}$ ergs/sec,
which may be larger than the power inferred from the giant radio
lobes ($\simless 10^{47}$ ergs/sec). While the former argument
may not strictly be applicable to the internal shock model (since
each knot is a separate entity and the distance from the source
is not a measure of the age of the source), the power requirement
for some sources  may indeed be very large, for e.g. 
Knot A of 1136-135 requires $P_{ave} \approx 2\times 10^{48}$ ergs/sec 
(Table 3). However, the energy requirement may be decreased if
the magnetic field is sub-equipartition e.g. 3C273 (Table 3).
Thus it is desirable to obtain direct observational signatures,
like spectral indices, to discriminate between models.

A realistic description of the knots will be more complicated than the
simple model considered here. For example, the forward and reverse
shocks that should form when the blobs collide, may provide 
different injection rates and at different
locations within the Knot. However, the physics of these shock formations
and the subsequent acceleration of particles is complicated and
unclear, especially if they are mediated by magnetic fields. In the
future, results from sophisticated numerical simulations may be
compared with higher resolution data ( which can resolve the
internal structure of the knots) to proof (or disproof) the
internal shock model.

In summary, a simple internal shock model is consistent
with the broad band spectra of knots in AGN kpc-scale jets.
The age of the knots ($t_{O,k}$) may be smaller than
the time-scale of injection of non-thermal particles ($T_{ON,K}$),
which implies there may be spectral  breaks in the 
synchrotron and IC spectra, instead of exponential high energy cutoffs.
The jets are powered by AGN, which sporadically eject out material on
a $10^{11-12}$ sec time-scale.  

\acknowledgements

The authors thank A. K. Kembhavi and C. L. Kaul for useful discussions.

\clearpage

\clearpage

\clearpage

\input psbox.tex


\begin{thebibliography}{}



\bibitem[Atoyan \& Dermer (2004)] {ATO04} Atoyan, A., and Dermer, C., 2004, \apj, {\it in press} (astro-ph:0402647).

\bibitem[Begelman \etal (1984)]{begel} Begelman, M. C., Blandford, R. D., Rees, M. J., 1984, Rev. Mod. Phys., 56, 255

\bibitem[Biretta \etal (1991)] {bire} Biretta, J. A., Stern, C. P., and Harris, D. E., 1991, \aj, 101,1632

\bibitem[Chang \& Cooper (1970)]{chang} Chang, J. S., and Cooper, G., 1970, Journal of Computational Physics, 6, 1

\bibitem[Chartas \etal (2000)] {chartas} Chartas, G. \etal, 2000, \apj, 542, 655

\bibitem[Dermer (1995)] {der} Dermer, C., 1995, \apj, 446, L63

\bibitem[Dermer \& Atoyan (2002)] {Der02} Dermer, C., and Atoyan, A., 1995, \apj, 568, L81

\bibitem[Jaffe \& Perola (1973)]{jaffe} Jaffe, W. J. \& Perola, G. C., 1973, \aap, 26, 412


\bibitem[Jester \etal (2005)]{Jes05} Jester, S., Rosser, H-J, Meisenheimer, K. \& Perley, R., 2005, \aap, 431, 477.

\bibitem[Heavens \& Meisenheimer (1987)]{heavens} Heavens, A., \& Meisenheimer, K., 1987, \mnras, 225, 335

\bibitem[Kardashev (1962)] {karda} Kardashev, N. S., 1962, Soviet Astron.-AJ, 6, 317

\bibitem[Kirk \etal (1998)] {kirk} Kirk, J. G., Reiger, F. M., Mastichiadis, A., 1998,\aap, 333,452

\bibitem[Lazzati \etal (1999)] {lazzati} Lazzati, D., Ghisellini, G., \& Celloti A.,1999, \mnras, 309, L13

\bibitem[Meisenheimer \etal (1989)] {meis} Meisenheimer, K. \etal, 1989, \aap, 219, 63

\bibitem[Marshall \etal (2001)] {marsh} Marshall, H. L. \etal, 2001, \apj, 549, 167

\bibitem[Pacholczyk (1970)] {pach} Pacholczyk, A. G., 1970, Radio Astrophysics (San Fransisco : Freeman)

\bibitem[Panaitescu \etal (1999)] {pana} Panaitescu, A., Spada, M., \& Meszaros, P., 1999, \apj, 522, L105

\bibitem[Rees (1978)] {rees78} Rees, M. J., 1978, \mnras, 184, 61

\bibitem[Rees \& Meszaros (1994)] {rees94} Rees, M. J., \& Meszaros, P., 1994, \apj, 430, L93

\bibitem[Sahayanathan \etal (2003)] {sahay} Sahayanathan, S., Misra, R., Khembhavi, A. K., and Kaul, C. L., 2003, \apj, 588, L77

\bibitem[Sambruna \etal (2001)] {samb01} Sambruna. R. M. \etal, 2001, \apj, 549, L161

\bibitem[Sambruna \etal (2002)] {samb02} Sambruna, R. M. \etal, 2002, \apj, 571, 206

\bibitem[Schwartz \etal (2000)] {schw} Schwartz, D. A. \etal, 2000, \apj, 540, L69

\bibitem[Spada \etal (2001)] {spa} Spada, M., Ghisellini, G., Lazzati, D., and Celotti, A., 2001, \mnras, 325, 1559

\bibitem[Perlman \etal (2001)] {perl} Perlman, E. S. \etal, 2001, \apj, 551, 206

\bibitem[Pesce \etal (2001)] {pesce} Pesce, J. E. \etal, 2001, \apj, 556, L79

\bibitem[Tavecchio \etal (2000)] {tav} Tavecchio, F., Maraschi, L., Sambruna, R. M., and Urry, C. M., 2000, \apj, 544, L23

\bibitem[Wilson \& Yang (2002)] {wilson} Wilson, A. S., and Yang, Y., 2002, \apj, 568, 133

\bibitem[Worall \etal (2001)]{worall} Worall, D. M., Birkinshaw, M., \& Hardcastle, M. J., 2001, \mnras, 326, L7
\end{thebibliography}
\end{document}